%% file: Main.tex
\documentclass[
    aps,
    prd,
    10pt,
    onecolumn,
    amsmath,
    amssymb,
    nofootinbib,
    longbibliography,
    showkeys
]{revtex4-2}

\usepackage[utf8]{inputenc}
\usepackage[T1]{fontenc}
\usepackage{microtype}

\usepackage{bm}
\usepackage{mathrsfs}
\usepackage{upgreek}

\usepackage[version=4]{mhchem}

\usepackage{siunitx}
\usepackage{graphicx}
\graphicspath{{Fig/}}

\usepackage{booktabs}
\usepackage{multirow}
\usepackage{dcolumn}
\usepackage[caption=false]{subfig}

\usepackage{xcolor}
\usepackage{comment}

\usepackage[
    colorlinks = true,
    linkcolor  = blue,
    citecolor  = blue,
    urlcolor   = blue
]{hyperref}

\hypersetup{
    pdftitle = {
        Spectroscopic study of argon electroluminescence light
        with a wavelength-sensitive particle detector
    },
    pdfauthor = {CIEMAT DM}
}

\begin{document}

\title{
Spectroscopic study of  argon electroluminescence light
with a wavelength-sensitive particle detector
}

\author{R. Santorelli}
\email{roberto.santorelli@ciemat.es}
\author{V. Pesudo Fortes}
\author{G. Vera Díaz}
\email{guillermo.vera@ciemat.es}
\author{C. K. Guérard}
\author{R. López Manzano}
\author{L. Luzzi}
\altaffiliation{Now at Department of Physics, University of California,
Davis, CA 95616, USA}
\author{J. Martínez Morales}
\author{L. Romero}
\affiliation{
CIEMAT,
Centro de Investigaciones Energéticas, Medioambientales y Tecnológicas,
Avenida Complutense 40,
28040 Madrid,
Spain
}

\date{\today}

\begin{abstract}
\input{Sec/Abstract}
\end{abstract}


\maketitle


\input{Sec/Introduction}
\input{Sec/Setup}

\input{Sec/Commissioning}
\input{Sec/Data1b}

\input{Sec/Data3b}

\input{Sec/Conclusions}

\input{Sec/Acknowledgments}





\bibliographystyle{apsrev4-2}
\bibliography{bibliografia}

\end{document}

%% file: Sec/Abstract.tex
We present a spectroscopic study of argon electroluminescence (EL) light in a gaseous time projection chamber (TPC). 
Using a compact detector equipped with photomultiplier tubes with
different spectral sensitivities, we measure the light emission in two
nominal wavelength regions, approximately [110, 160]~nm and
[160, 650]~nm, at different gas pressures.
In addition to the well-known 128 nm emission of the second continuum, significant emission is observed in the [160, 650] nm band, with a prompt, nanosecond-scale response indicating that photon production closely follows the transit of drifting electrons across the high-field EL region.
In contrast, the [110, 160] nm emission displays a markedly slower time evolution, dominated by excimer formation and de-excitation dynamics, with the light output governed primarily by the long-lived triplet component of the argon second continuum emission.
We demonstrate that the full VUV pulse shape can be reproduced by convolving
the fast UV3 response with excimer formation and decay functions, providing a
coherent phenomenological interpretation of the observed signals. Our results provide new insights into the spectral and temporal properties of argon electroluminescence and have direct implications for the design and optimization of next-generation rare-event detectors based on gaseous TPCs. Further studies are underway to characterize this emission more precisely and to investigate its potential for particle discrimination.

%% file: Sec/Introduction.tex
\section{Introduction}
\label{sec:Intro}

Time projection chambers (TPCs) based on noble elements are widely employed in rare-event searches due to their scalability and their ability to provide three-dimensional event reconstruction through the combined measurement of ionization and scintillation signals \cite{ArDM:2017ndf,DarkSide:2018bpj,XENON:2024ijk}. In a typical dual-phase (liquid–gas) configuration, primary scintillation photons are produced in the liquid target, while the ionization charge is extracted into a thin gas layer, where electrons are accelerated by a high electric field operated below the onset of Townsend multiplication. In this proportional regime, excitation processes dominate over secondary ionization and charge multiplication is suppressed, resulting in the production of electroluminescence (EL) photons with intrinsically low statistical fluctuations \cite{DarkSide:2018ppu}.
This secondary light signal is proportional to the charge extracted from the liquid, with a photon yield that scales with both the electron drift path in the gas and the reduced electric field (E/P), providing a stable and controllable optical gain. In gaseous argon, typical reduced electric fields for EL operation range from $2$ to $5~\mathrm{kV\,cm^{-1}\,bar^{-1}}$, and the reduced EL yield has been
experimentally parameterized as a linear function of the reduced electric field,
with a proportionality coefficient of order
$100~\mathrm{photons\,electron^{-1}\,kV^{-1}}$~\cite{Monteiro2008}.

This linear photon production along the electron trajectory provides a readily measurable light signal, enabling operation in large-scale detectors and sensitivity down to single electrons extracted from the liquid \cite{DarkSide:2018ppu}. Such sensitivity is crucial for a broad range of experiments requiring the detection of very small ionization signals, such as searches for low-mass dark matter and coherent elastic neutrino--nucleus scattering (CE$\nu$NS). Since both the primary scintillation (S1) and the secondary electroluminescence signal (S2) are recorded by the same photon sensors, the detector design remains relatively simple while retaining precise timing, spatial reconstruction, and interaction multiplicity information.

The dominant contribution to photon emission in noble gases is commonly attributed to the so-called second continuum, which in argon consists of an emission band a few nanometers wide, centered around 128~nm, and arises from the radiative decay of argon excimers to the dissociative ground state~\cite{Santorelli:2020fxn}. The assumption of quasi-monochromatic vacuum-ultraviolet emission has historically guided the design of noble-gas detectors.

In most argon and xenon EL-TPCs, light-detection systems rely on broad-band optical sensors to record both primary scintillation and secondary EL signals, often combined with wavelength shifters. This approach integrates the light over a wide spectral range and does not exploit the possible composite nature of the emission spectrum.

Recent experimental findings have revealed additional components in the scintillation spectrum of noble elements \cite{Santorelli:2020fxn,Leardini:2021qnf}. In particular, a fast emission band extending between 160 and 300 nm has been observed in argon. This component exhibits a nanosecond-scale time structure and represents the dominant contribution to the prompt scintillation signal in gaseous argon.
Its origin has been commonly associated with the so-called third continuum, although alternative mechanisms cannot be excluded. Regardless of its microscopic nature, these observations clearly demonstrate that argon scintillation is intrinsically multi-component, both spectrally and temporally, with distinct emission channels governed by different physical processes. In our previous work, we reported, to our knowledge, the first systematic wavelength- and time-resolved study of gaseous-argon scintillation induced
by individual $\alpha$ and $\beta$ sources, and explicitly introduced
scintillation spectroscopy as a novel route to particle
identification~\cite{Santorelli:2020fxn}. The present study extends this
program to electroluminescence, establishing a common wavelength-resolved
framework for the primary and secondary light signals produced in gaseous
argon TPCs.

In this context, the spectral properties of argon electroluminescence in the ultraviolet–visible (UV–VIS) region remain largely unexplored compared to the extensively studied vacuum ultraviolet (VUV) range. Establishing the presence of emission at wavelengths beyond the VUV and clarifying the underlying mechanisms are important for the design of next-generation rare-event detectors.
The spectral composition of EL light directly impacts photon detection efficiency, light propagation in the active volume, and the accuracy of event reconstruction. A quantitative characterization of these longer-wavelength components is therefore necessary to fully exploit the information carried by the argon electroluminescence signal.

In this work, we present a spectroscopic characterization of argon electroluminescence using a compact TPC equipped with photomultiplier tubes sensitive to distinct wavelength ranges. By comparing the signals recorded in the nominal VUV (110--160~nm)
and UV--VIS (160--650~nm) sensitivity regions, we quantify the relative
contribution of the different spectral components.
We measure a UV3-to-UV2 photon-yield ratio of approximately 10\% for the
component temporally correlated with electroluminescence at both pressures.
At 3~bar, integration of the full UV3 waveform, including an additional early
component whose origin remains under investigation, yields a larger ratio of
approximately 20\%.
These results provide new insights into the spectral structure of argon electroluminescence in particle detectors and highlight the importance of a wavelength-resolved description of the emission. Such a description can improve detector optical models and enable new analysis strategies that exploit spectral information to enhance detector performance.

%% file: Sec/Setup.tex
\section{Experimental Setup}
\label{sec:Setup}

The experimental setup is designed to perform a spectrally resolved study
of electroluminescence (EL) light in gaseous argon (GAr) using a compact TPC
equipped with two PMT systems with complementary wavelength-dependent spectral
responses, predominantly probing the emission below and above 160~nm,
respectively.

\begin{figure}[!th]
      \centering
        \begin{tabular}{c}
            \includegraphics[width=0.4\textwidth]{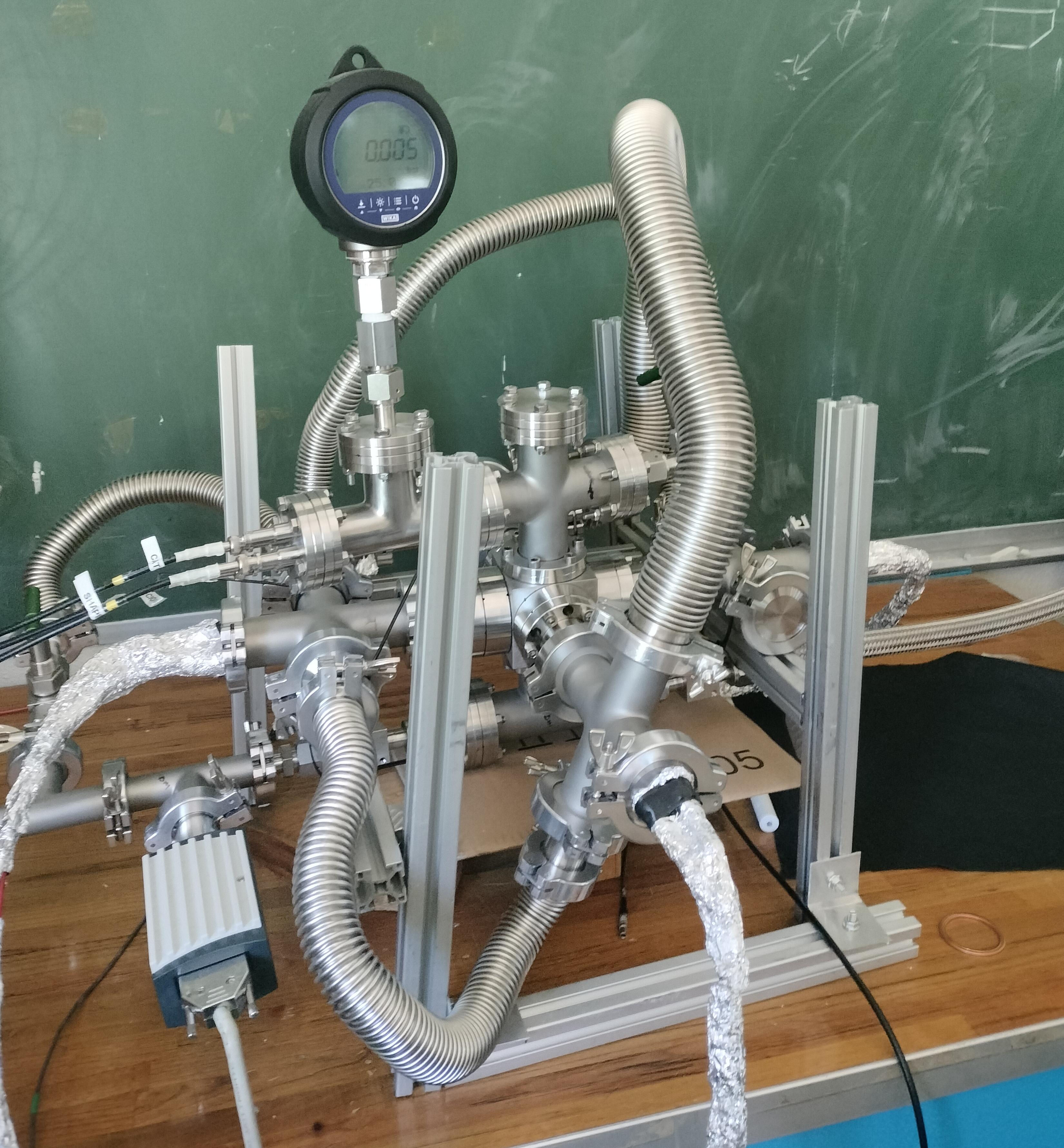}
            \small(a)
        \end{tabular}
        \begin{tabular}{c}
            \includegraphics[width=0.4\textwidth]{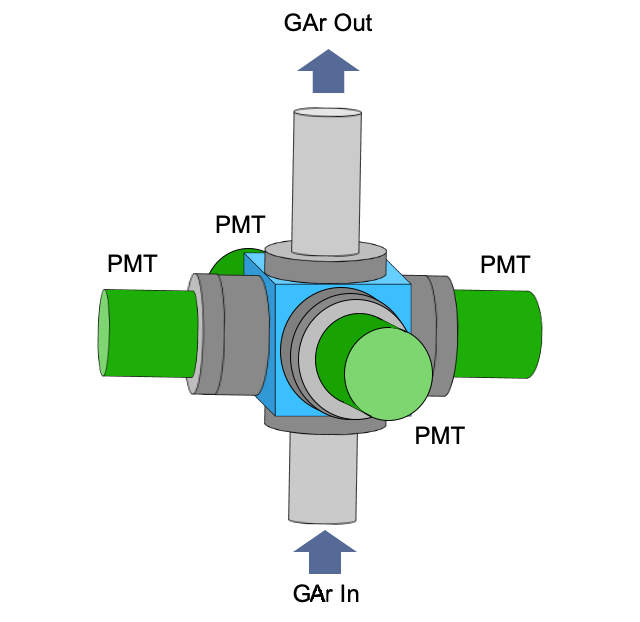}
            \small(b)
        \end{tabular}
        \caption{Picture (a) and schematic (b) of the experimental setup.}
        \label{fig:hpgas}
\end{figure}

The chamber, shown in Fig.~\ref{fig:hpgas}, is a stainless steel cube with a side length of 7 cm, equipped with six CF-40 flanges, one on each face. Four of these flanges accommodate MgF$_2$ optical viewports, designed by the CIEMAT Dark Matter group and custom-fabricated in-house. These windows allow light transmission for wavelengths above $\lambda \gtrsim 110$ nm, while the remaining two flanges serve as operational connections. The MgF$_2$ viewports have a transmission coefficient of approximately 95\% for wavelengths above 180 nm and about 33\% at 128 nm \cite{Santorelli:2020fxn}.

The top and bottom flanges are connected to the vacuum and gas handling systems, while the lateral flanges house four PMTs mounted in light-tight cylindrical enclosures, where a vacuum (P $< 5 \times 10^{-4}$ mbar) is maintained during operation. Each enclosure is equipped with a KF-40 4-pin electrical feedthrough providing connections for voltage bias and signal readout.

To ensure adequate GAr purity, the chamber is evacuated prior to filling, reaching a pressure below $9 \times 10^{-5}$ mbar. The gas used in this experiment was Alphagaz\texttrademark-2, supplied by Air Liquide, with a nominal purity of 99.9999\% and O$_2$ and H$_2$O concentrations at the ppm level. Under these conditions, their impact on the spectral distribution of the emitted light is considered negligible. 
A continuous argon flow from bottom to top is maintained during data taking to preserve gas purity and minimize contamination from outgassing of internal surfaces. The chamber pressure, in the range 1--3 bar, is monitored and controlled using a WIKA\texttrademark\ CPG1200 pressure gauge.

Four PMTs, covering two complementary wavelength regions, are mounted on the
lateral flanges of the chamber. The effective spectral response of each channel
is determined by the wavelength-dependent PMT quantum efficiency and by the
transmission of the optical elements along the light path. The nominal
sensitivity intervals are summarized in Table~\ref{tab:Rangedef} and should not
be interpreted as sharp bandpass boundaries. The narrower intervals reported
in our previous scintillation study~\cite{Santorelli:2020fxn} were constrained
using dedicated measurements with additional optical filters, whereas the
intervals adopted here describe the nominal response of the unfiltered PMT
channels.

Two VUV-sensitive PMTs, equipped with CsI photocathodes and MgF$_2$ windows
(Hamamatsu R6835\footnote{Datasheet available at:
\url{https://www.hamamatsu.com/content/dam/hamamatsu-photonics/sites/documents/99_SALES_LIBRARY/etd/R6835_TPMH1263E.pdf}}),
are mainly sensitive in the range [110, 160]~nm, hereafter referred to as the
UV2 region. The other two PMTs, designed for UV--VIS detection, feature bialkali
photocathodes and fused-silica windows
(Hamamatsu R7378\footnote{Datasheet available at:
\url{https://www.hamamatsu.com/content/dam/hamamatsu-photonics/sites/documents/99_SALES_LIBRARY/etd/R7378A_TPMH1288E.pdf}}),
and are mainly sensitive in the range [160, 650]~nm, hereafter referred to as
the UV3 region. Together, the two PMT systems allow simultaneous
wavelength-resolved detection of light predominantly below and above 160~nm.

\begin{table*}[ht!]
\centering
\caption{Definition of the nominal UV2 and UV3 detector-response regions.
The quoted intervals indicate the main spectral sensitivity of the PMTs and
do not represent sharp bandpass boundaries.}
\begin{tabular}{ c | c | c }
\hline
Region & PMT model & Range\\ 
\hline 
\hline 
\rule{0pt}{3.0ex}
UV2  & R6835 & [110,~160] nm\\ 
\rule{0pt}{3.0ex}
UV3 & R7378 & [160,~650] nm   \\ 
\hline  \hline
\end{tabular}
\label{tab:Rangedef}
\end{table*}

The PMTs of the same type are placed opposite each other, ensuring symmetric coverage and improving detection efficiency. At representative reference wavelengths, the quantum efficiencies are comparable, being approximately 0.15 at 128 nm for the UV2 PMTs and 0.18 at 200 nm for the UV3 PMTs \cite{Santorelli:2020fxn}. The UV3 PMTs were calibrated using their single-photoelectron response measured
in vacuum, while the UV2 PMTs were calibrated using single-photoelectron pulses
from argon scintillation.

The PMT signals are digitized using a CAEN DT5730SB ADC with 14-bit resolution, a sampling rate of 500 MS/s, and a 2 V dynamic range. The data are processed with custom software based on CAEN libraries, from which the relevant analysis variables are extracted. The trigger requires a coincidence between at least one pair of PMTs of the same type. The acquisition window is set to 60~$\upmu$s to capture the full EL signal, with 20\% of the window allocated to the pre-trigger region for baseline recording.

A more detailed description of the general setup can be found in \cite{Santorelli:2020fxn}. In the following, we describe the specific modifications implemented to generate and study EL light.

\begin{figure}[!ht]
      \centering
        \begin{tabular}{c}
            \includegraphics[width=0.29\textwidth]{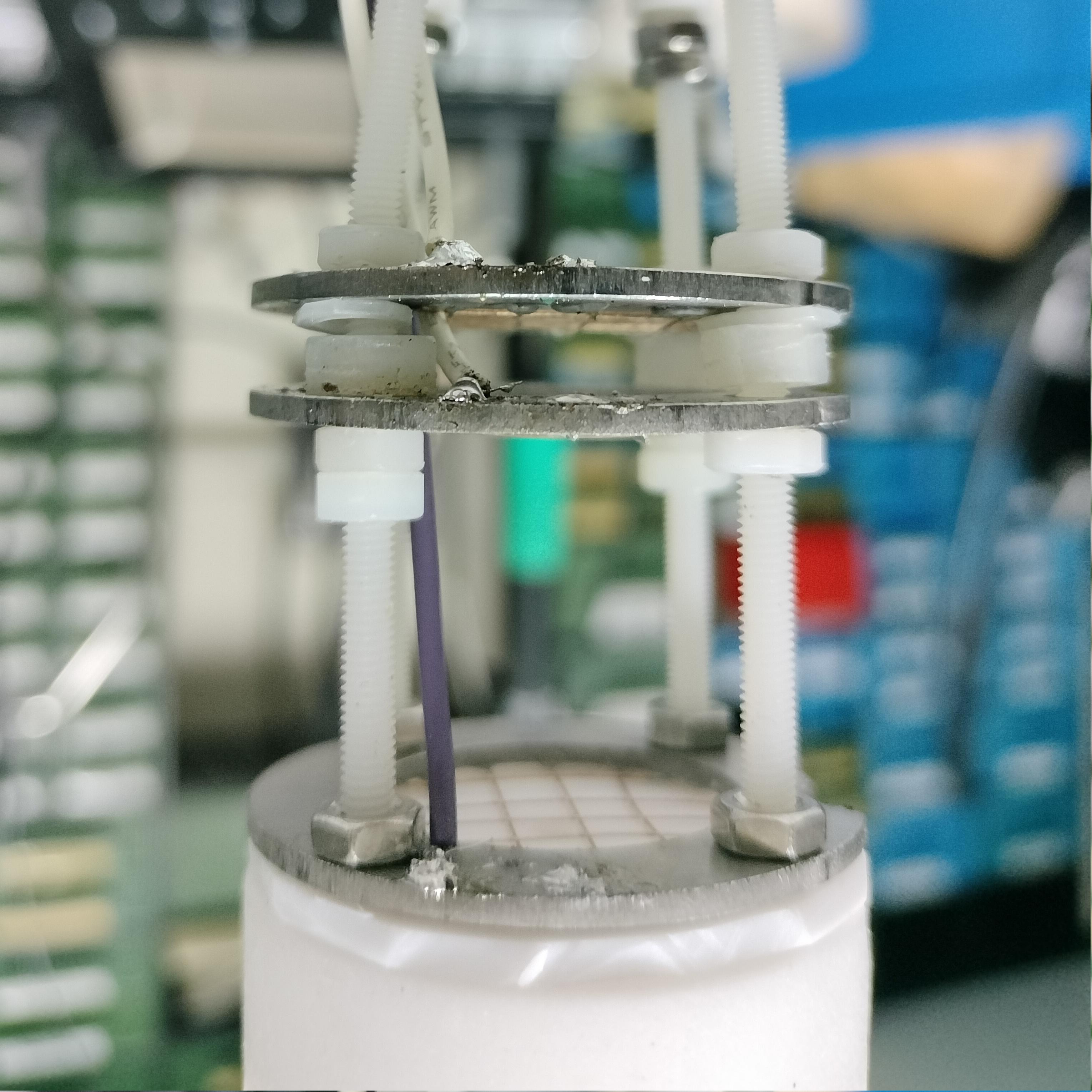}
        \end{tabular}
        \begin{tabular}{c}
            \includegraphics[width=0.33\textwidth]{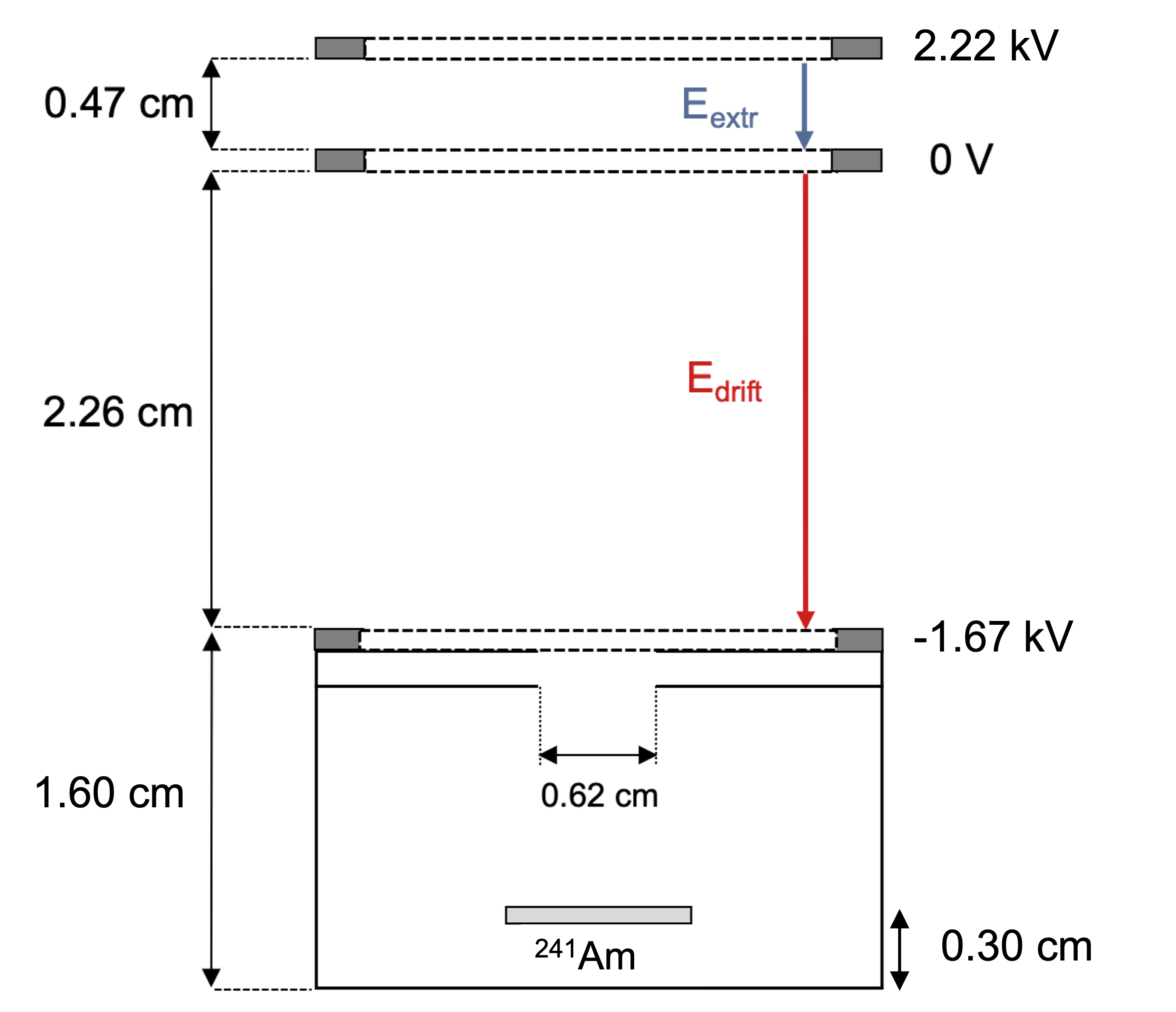}
        \end{tabular}
        \begin{tabular}{c}
            \includegraphics[width=0.29\textwidth]{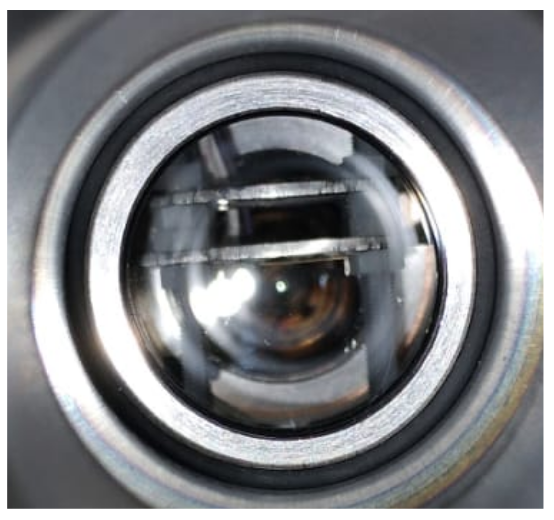}
        \end{tabular}
        \caption{Picture (left) and schematic (middle) of the TPC structure inside the chamber. Side view of the TPC from an optical MgF$_2$ viewport (right), showing the two metal frames of the grids that limit the EL region.}
        \label{fig:rejillas}
\end{figure}

Inside the chamber, a TPC structure specifically designed for EL spectroscopy is installed, as shown in Fig.~\ref{fig:rejillas}-left. The TPC consists of three ring-shaped aluminum disks supporting wire grids held at different voltages to establish the required electric fields. The disks provide mechanical support while leaving the central region open to allow the transmission of electrons and photons.

The drift region (Fig.~\ref{fig:rejillas}-middle), defined between the bottom and middle grids, operates at an electric field of $E_\text{drift} = 266$ V/cm, driving ionization electrons toward the EL region. Above the middle grid, a higher electric field of $E_\text{EL} = 1.69$ kV/cm is applied between the middle and top grids, accelerating the electrons and producing EL light. Below the bottom grid, a collimated $^{241}$Am $\alpha$ source ($\sim$ 5.5 MeV, $\sim$ 90 Bq) ionizes the argon gas.

The source is housed in a Teflon cylinder with a small opening (0.62 cm, Fig.~\ref{fig:rejillas}-middle), providing a nearly vertical collimation of $\mathrm{
\alpha}$ particles prior to the software-based cuts described in Sec.~\ref{sec:Comm}.

This configuration enables stable EL signal collection and provides the basis for the spectrally resolved analysis of EL light presented here.

%% file: Sec/Commissioning.tex
\section{Study of scintillation light and EL event selection at 1 bar}
\label{sec:Comm}

The first campaign was performed with pure Ar (99.9999\% or higher) at a
pressure of $1.08 \pm 0.05$~bar, hereafter referred to as the 1-bar data set. Data-taking runs of approximately 15 minutes were performed during chamber commissioning to prevent degradation of the apparatus. This approach reduces the probability of electrical discharges that could damage the detector components. Measurements were performed in three configurations: no fields (NF), drift field only (DF), and both drift and electroluminescence fields active (BF). A preliminary commissioning of the chamber, focused on studying scintillation light, was carried out using the NF and DF configurations.  

\begin{figure}[!ht]
     \centering
        \begin{tabular}{c}
         \includegraphics[width=0.47\textwidth]{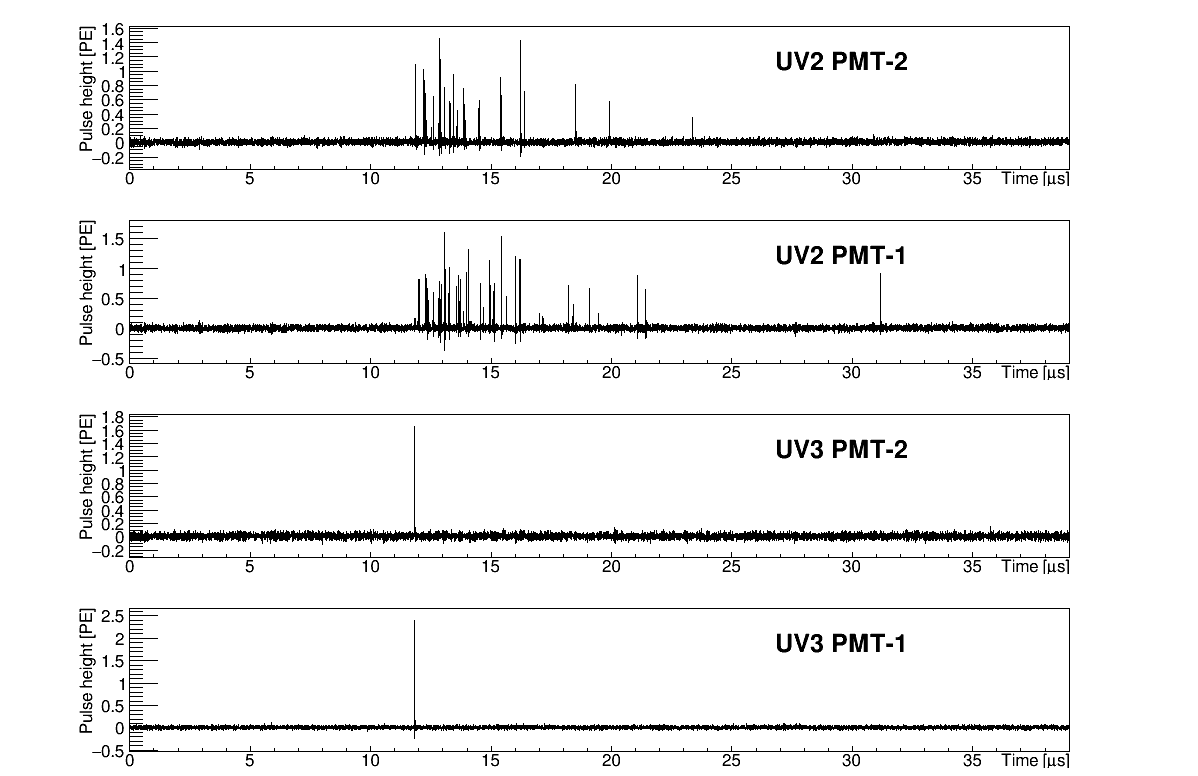}
        \end{tabular}
        \begin{tabular}{c}
         \includegraphics[width=0.47\textwidth]{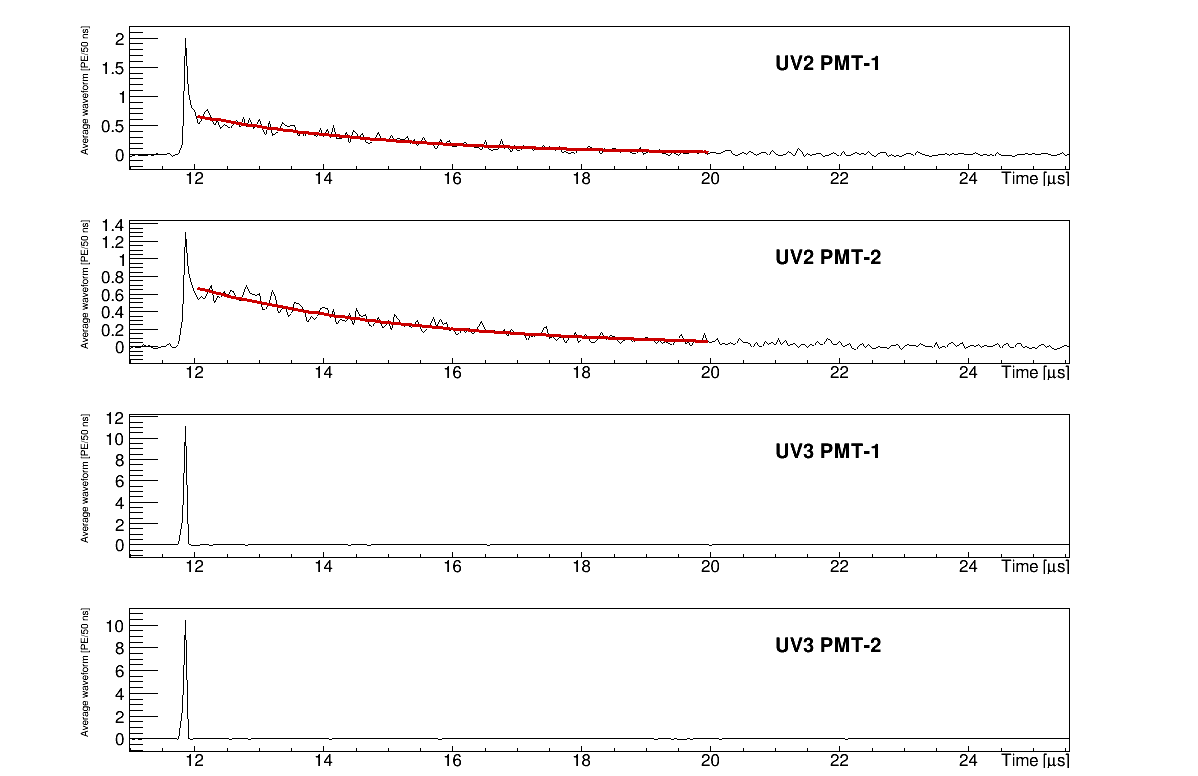}
        \end{tabular}
    
    \caption{Example waveform for the NF configuration (left) and average waveforms
at 1.08~bar for each PMT model in the NF configuration (right). An exponential fit is performed on the UV2 PMT channels.}
     \label{fig:NF_wave}
\end{figure}

The primary goal of the NF analysis is to characterize the primary scintillation emission generated by $\alpha$ interactions in GAr and to study its spectral and temporal properties in the two wavelength regions. Figure~\ref{fig:NF_wave}-left shows representative waveforms of $\alpha$-induced scintillation events recorded by the four PMTs under the operating conditions described above. A pronounced difference in temporal structure between the two spectral bands is clearly visible: the UV3 channel exhibits a fast component on the nanosecond timescale,
whereas the UV2 PMTs record a much broader emission extending over several microseconds. These distinct temporal features are consistent with previous measurements in gaseous argon~\cite{Santorelli:2020fxn}, where the fast component observed in the UV3 region was associated with the third continuum emission, while the long-lived emission detected in the UV2 band was attributed to the second continuum.

Figure~\ref{fig:NF_wave}-right shows the time profile of the scintillation produced by $\alpha$ interactions, averaged over 100 events and normalized to the peak value of the prompt scintillation signal in UV3 PMT-2. Exponential fits to the slow component of the argon scintillation detected by the UV2 PMTs yield triplet decay time constants of $\uptau_{\mathrm{PMT1}} = 3.11 \pm 0.11~\upmu\mathrm{s}$ and $\uptau_{\mathrm{PMT2}} = 3.19 \pm 0.15~\upmu\mathrm{s}$, where the quoted uncertainties include both the statistical error of the fit and the systematic uncertainty associated with the choice of the fit interval. 
These values are consistent with previous measurements performed with the same chamber operated without the field cage, as well as with the expected triplet lifetime in gaseous argon at ppm-level residual impurities, as specified for the Alphagaz\texttrademark-2 used in this study.

In the DF configuration, the overall waveform structure remains essentially unchanged, with the dominant contribution given by the primary scintillation signal. A small delayed component is nevertheless observed, particularly in the UV3 channels, where a few photons are detected after the prompt peak. This emission is attributed to distortions of the drift field near the external stainless-steel ring, where some drift lines may not terminate on the middle grid but instead reach the chamber walls, producing localized and unintended electroluminescence. A minor contribution from electron collection at metallic surfaces inside the chamber cannot be excluded. For the purposes of the present study, this effect is negligible. The associated light yield is significantly smaller than that observed in the EL configuration, and in the EL analysis the average DF waveform is subtracted from the signal, ensuring that the measured EL light is not contaminated by primary $\alpha$ scintillation or by this small drift-induced contribution.

Data in the BF configuration (i.e. with both drift and EL fields on) at 1 bar were collected by setting the voltages in the field cage as described in Sec.~\ref{sec:Setup}. The maximum electron transit times across the drift region (2.26~cm, Fig.~\ref{fig:rejillas}-middle) and the EL region (0.47~cm) are approximately $7.9 \pm 0.8~\upmu$s and $1.2 \pm 0.1~\upmu$s, respectively, for electrons drifting along vertical trajectories. These values are obtained by solving the electrostatic field configuration with appropriate boundary conditions on the grids using the PDE solver provided by MATLAB \cite{MATLAB}. The solution shows non-uniform electric field intensities due to the discrete structure of the wire grids. The drift time is estimated using the electric field at the center of the TPC and the corresponding drift velocity from~\cite{nakamura1988electron}. The quoted uncertainties reflect the variations in the electric field obtained from the field calculations.

At 1~bar, 5.5~MeV $\alpha$ particles from the $^{241}$Am source have a range of approximately 4.4~cm. As a result, $\alpha$ particles emitted vertically traverse the EL region, leading to a temporal overlap between the EL signal and the primary scintillation. In practice, this does not affect the analysis, since the EL light yield is much larger than that of the primary scintillation.

Although the source activity is relatively low (approximately 45~$\alpha$ particles per second emitted into the argon), and the collimator further reduces the event rate by selecting nearly vertical tracks, data-taking runs of a few tens of minutes under stable conditions are sufficient to collect a statistically significant sample of EL events (several hundred).

\begin{figure}[!ht]
     \centering
    \begin{tabular}{c}
         \includegraphics[width=\textwidth]{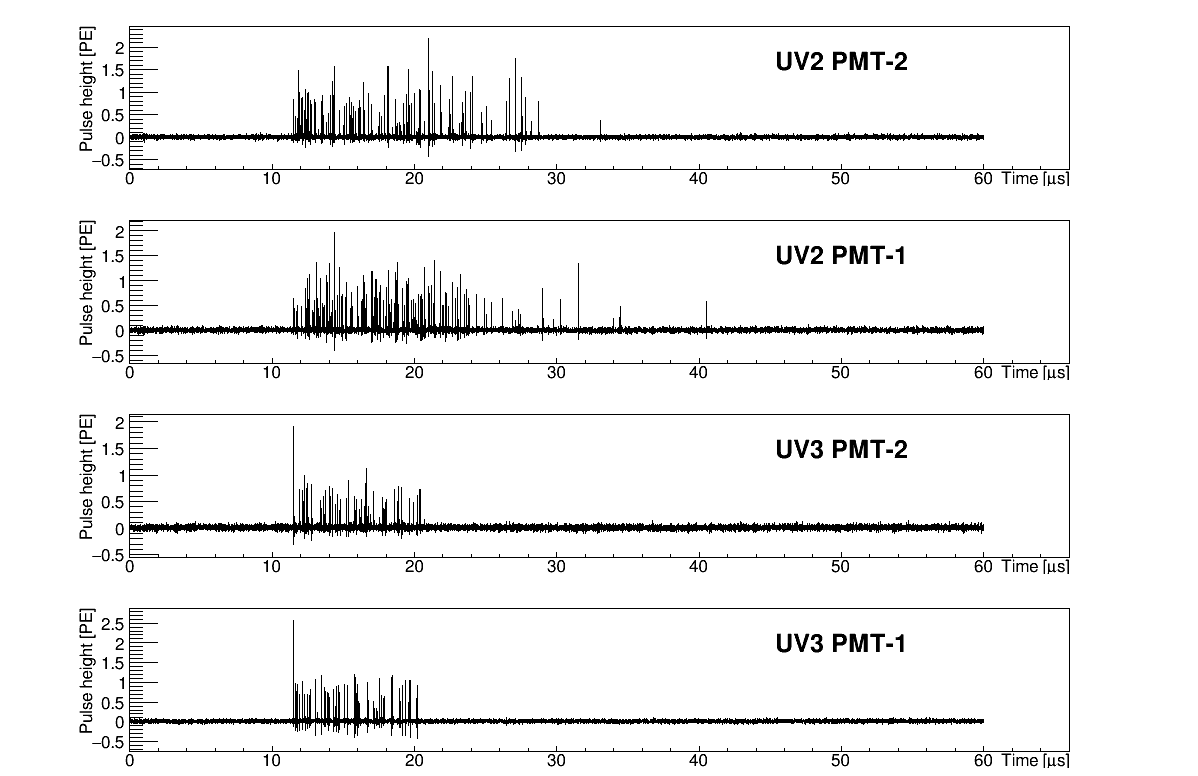}
    \end{tabular}
    \caption{Example of a raw waveform of the BF configuration.}
     \label{fig:waveforms_EL}
\end{figure}

Figure~\ref{fig:waveforms_EL} shows a representative raw waveform of an $\alpha$ interaction in argon recorded by the four PMTs in the BF configuration. With the EL field on, a strong emission delayed with respect to the prompt scintillation is observed in both spectral regions. This emission disappears when the EL field is turned off and is therefore identified as electroluminescence light produced by ionization electrons drifting into the high-field region.

The presence of EL light in both channels indicates that part of the emission
lies beyond the VUV range, within the nominal spectral sensitivity region of
the UV3 PMTs (160--650~nm). The delayed signal observed in the UV2 channel dominates the waveform and is significantly larger than the primary scintillation emission. The comparison between the two bands shows that the temporal structure of the EL signal differs, with the UV3 component evolving faster than the UV2 emission.

For the quantitative analysis of EL signals, a set of software selection cuts was applied to obtain a clean and well-defined sample of $\alpha$ interactions occurring in the central region of the chamber. The selection strategy consisted of a sequence of low-level cuts designed to ensure waveform stability and the presence of a genuine electron-induced luminescence. Events characterized by high electronic noise were suppressed by rejecting signals with large RMS values in the pre-trigger region, while baseline fluctuations were controlled by requiring consistency of the average signal
level between the pre-trigger region and the end of the waveform.

Pile-up and mis-triggered events were rejected by constraining the position of the prompt scintillation peak in the UV3 channels to the expected trigger window. Finally, events not producing EL light were removed by imposing a minimum threshold on the integrated charge in the EL time window.

In addition, a high-level geometrical selection was implemented through a symmetry requirement between pairs of opposite PMTs. The asymmetry parameter is defined as
\[
A_{UVi} = \frac{Q_{L}^{\mathrm{PMT1}} - Q_{L}^{\mathrm{PMT2}}}
{Q_{L}^{\mathrm{PMT1}} + Q_{L}^{\mathrm{PMT2}}},
\]
where $Q_L$ is the integrated charge in the EL time window, defined from 200~ns after the trigger and extending over 28~$\upmu$s, and PMT1 and PMT2 are opposite sensors of the same type. Events were required to satisfy $\left|A_{\mathrm{UV3}}\right| < 0.15$ and $\left|A_{\mathrm{UV2}}\right| < 0.20$. This asymmetry cut further constrains the event topology beyond the hardware collimation, selecting nearly vertical $\alpha$ tracks and ensuring that the ionization electrons drift along the detector axis. Consequently, the EL light is produced in the central region of the gap.

These selections efficiently reject noisy events, typically characterized by long trains of pulses associated with sparks or corona effects between the field rings and the chamber walls, as well as $\alpha$ particles emitted at large angles, which either do not enter the EL region or generate EL close to the boundaries of the high-field region. The resulting sample is therefore composed of well-centered EL events suitable for spectral analysis.
Approximately 280 events survive the cuts out of $2 \times 10^4$
triggers, corresponding to about 1.4\% of the total.

%% file: Sec/Data1b.tex
\section{Electroluminescence signal spectral analysis at 1~bar}
\label{sec:Data1b}

\begin{figure}[!ht]
    \centering
    \includegraphics[width=\textwidth]{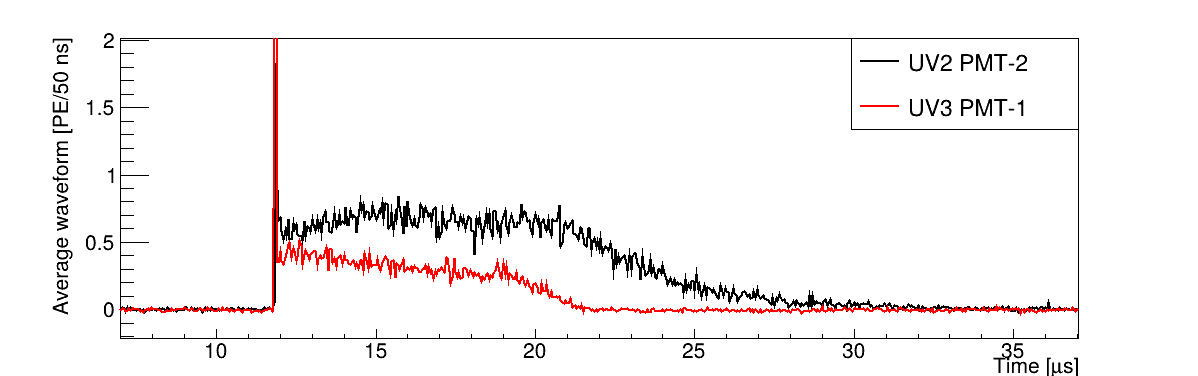}
 \caption{Averaged waveforms for two PMTs of different types, representative of the UV2 and UV3 spectral components.}
    \label{fig:av_waveform}
\end{figure}

An averaged waveform of the events surviving the selection cuts is shown in Figure~\ref{fig:av_waveform} for two representative PMTs of different types, while the behavior of the other PMTs of the same type is similar. The signal is resampled every 25 samples (50 ns). A clear difference in temporal structure between the UV3 and UV2 signals is observed. The earlier return of the UV3 waveform to baseline compared to UV2 indicates that it is associated with a faster emission process. 
In particular, the exponential behavior observed in the UV2 PMTs starts when the UV3 signal has vanished and is consistent with the decay of the triplet excimer state, once the last electrons have crossed the EL region.

\begin{figure}[!ht]
    \centering
    \includegraphics[width=\textwidth]{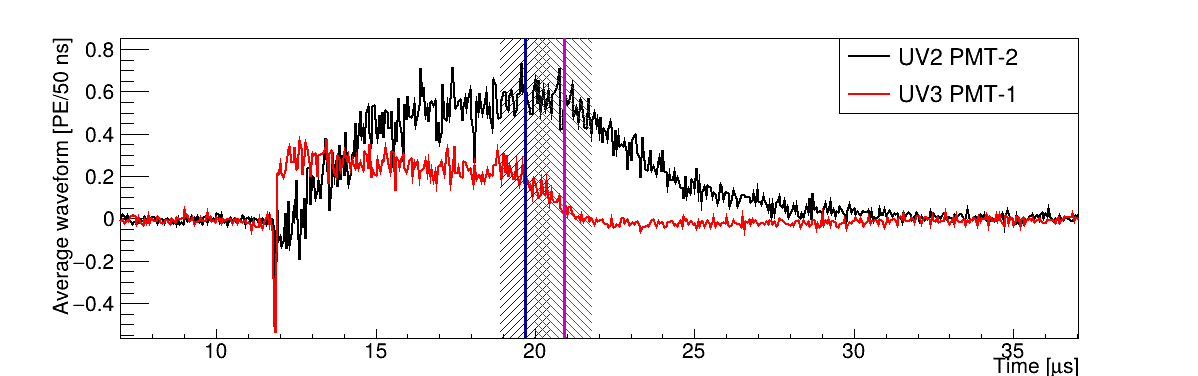}
    \caption{Averaged waveforms in the BF configuration after subtraction of the corresponding DF signals for UV2 PMT-2 and UV3 PMT-1. The shaded regions indicate the maximum expected drift time, $7.9 \pm 0.8$~$\upmu$s (blue), and the maximum time to cross the EL region, $1.2 \pm 0.1$~$\upmu$s (purple).}
    \label{fig:av_waveform_subs}
\end{figure}

In Figure~\ref{fig:av_waveform_subs}, the waveforms are shown after subtraction of the corresponding signals obtained in the DF configuration (EL turned off), removing the contribution from primary scintillation light and isolating the EL component. The figure shows that the signal in the UV3 channel develops almost simultaneously with the drift of the electrons through the EL region, as expected from the extended ionization track produced by the $\alpha$ particle across the chamber.

The UV3 signal initially exhibits a nearly linear behavior with a small negative slope during the first $\approx 7~\upmu$s, showing a decrease of roughly 21\%. This behavior can be due to the balance between electrons entering and leaving the EL region and is consistent with the maximum drift time of electrons produced at the beginning of the ionization track near the cathode, as indicated by the blue shaded region in the figure. The observed slope can be attributed to the variation of the ionization density along the $\alpha$ track and is consistent, at first order, with the expected Bragg curve for 5.5~MeV $\alpha$ particles, which reflects the increase in ionization density as the particle loses energy along its path through the chamber.

At $\approx 7~\upmu$s after the trigger, a change in the slope of the signal is observed. This corresponds to the time at which electrons produced near the cathode (i.e. at the beginning of the track) have completed their drift and reached the EL region. From this point on, no additional ionization electrons enter the EL gap, and the signal is produced only by those already drifting through the high-field region. Finally, the signal vanishes within the expected maximum electron transit time across the full drift and EL regions, $9.1 \pm 0.9$~$\upmu$s, as indicated by the purple shaded region in the figure.

In contrast to the fast UV3 emission, the averaged waveform in the UV2 spectral region does not rise immediately after the trigger, but exhibits a delayed onset before increasing. This behavior can be interpreted as a consequence of the second continuum photon production triggered by electrons accelerated in the high-field region. Due to its fast photon emission mechanism, the UV3 emission closely follows the time evolution of the electrons drifting in the chamber, and its temporal structure is therefore directly governed by the electron dynamics. In contrast, the UV2 emission reflects the combined effect of charge transport and the formation and de-excitation times of argon excimers, with the latter occurring on timescales comparable to or longer than the time
required for the electrons to cross the EL region in GAr at 1~bar.

We estimate the ratio between the electroluminescence photons emitted in the UV3 and UV2 spectral ranges by integrating the waveforms shown in Figure~\ref{fig:av_waveform_subs}, obtaining the number of photoelectrons detected by each type of PMT during the electroluminescence emission time. The result is then corrected for the quantum efficiency (QE) of the PMTs and for the transmission of the MgF$_2$ windows at the corresponding wavelengths, yielding a UV3/UV2 ratio of $9.8 \pm 2.9$\%. This result shows that the UV3 contribution to the electroluminescence light
is non-negligible, representing a sizable contribution relative to the UV2
emission.

Given that the fast UV3 emission directly follows the electron drift, while the UV2 signal is shaped by the formation and de-excitation of the argon excimers involved, the correlation between the emission in the two spectral regions can be investigated. To this end, the UV2 waveform is modeled as
\begin{equation}\label{eq:UV3_conv}
    U_{UV2}(t) = A \left[U_{UV3} \otimes I_{UV2}\right](t-\Delta t)
\end{equation}

where the function $I_{UV2}(t)$, describing the temporal response of the second continuum excimer emission~\cite{Santorelli:2020fxn}, is defined as
\begin{equation}\label{eq:h_2}
    I_{UV2}(t)=f_s\dfrac{e^{-t/\tau_s}-e^{-t/\tau_f}}{\tau_s-\tau_f}+f_t\dfrac{e^{-t/\tau_t}-e^{-t/\tau_f}}{\tau_t-\tau_f}.
\end{equation}

Here, $f_s$ and $f_t$ denote the relative fractions of light emitted in the singlet and triplet channels, respectively, while $\tau_s$ and $\tau_t$ are the singlet and triplet decay times, and $\tau_f$ is the excimer formation time.

The UV2 waveform is fitted with the convolution model defined in Eq.~\eqref{eq:UV3_conv}, using the measured UV3 signal as input. The fit yields a value of $f_s$ compatible with zero, indicating that the emission is dominated by the triplet component for EL-induced signals. The extracted values of the excimer formation time $\tau_f$ and decay times are consistent with those reported in the literature at 1~bar~\cite{Santorelli:2020fxn}. 

The ability to reproduce the UV2 pulse shape from the UV3 waveform demonstrates a clear correlation between the two spectral components and supports the interpretation that the UV3 emission is directly associated with the electron drift, while the UV2 signal arises from the subsequent formation and de-excitation of argon excimers.

\begin{figure}[!ht]
    \centering
    \includegraphics[width=0.9\textwidth]{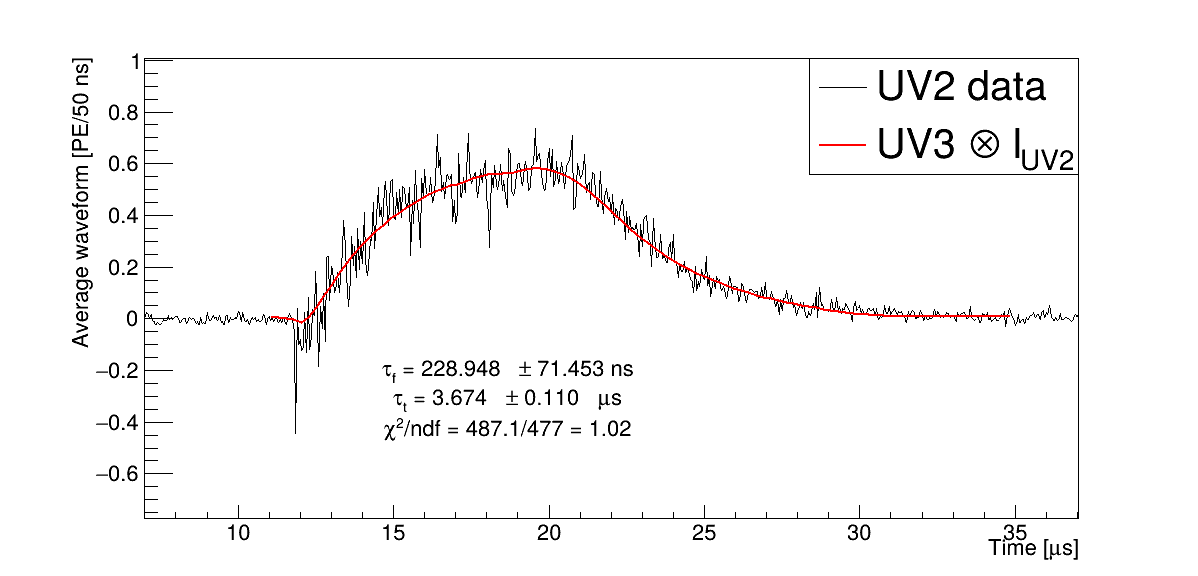}
    \caption{Fit of the UV2 waveform using the convolution model defined in Eq.~\eqref{eq:UV3_conv}. The convolution of the UV3 signal with the excimer formation and de-excitation response can reproduce the observed UV2 waveform.}
    \label{fig:UV2_fit}
\end{figure}

%% file: Sec/Data3b.tex
\section{Preliminary Study of the Electroluminescence Signal at 3~bar}
\label{sec:Data3b}

In order to study the behavior of electroluminescence light at higher pressure
under comparable conditions, data were collected at 3~bar while preserving the
same reduced drift and EL fields, $E_{\mathrm{drift}}/P$ and
$E_{\mathrm{EL}}/P$, as in the 1.08~bar measurements. To this end, the electric
fields were set to $E_{\mathrm{drift}} = 738~\mathrm{V\,cm^{-1}}$ and
$E_{\mathrm{EL}} = 4.72~\mathrm{kV\,cm^{-1}}$.

This configuration is of particular interest, since the gas density of argon at 3~bar and 300~K  is comparable to that of argon at 1~bar and 87~K ($\approx~5.5\times10^{-3}$~g/cm$^{3}$), as in the gas phase of dual-phase liquid argon TPCs based on electroluminescence.

At this pressure, the range of 5.5~MeV $\alpha$ particles is significantly reduced ($\approx 1.45$~cm). As a consequence, unlike the 1~bar case, the $\alpha$ tracks do not traverse the full chamber, but stop just above the cathode, at the beginning of the drift region, producing a localized ionization distribution (Fig.~\ref{fig:rejillas}). The same selection cuts used in the 1~bar analysis (Sec.~\ref{sec:Comm}) were applied to select nearly vertical ionization tracks.

\begin{figure}[!t]
    \centering
    \includegraphics[width=\textwidth]{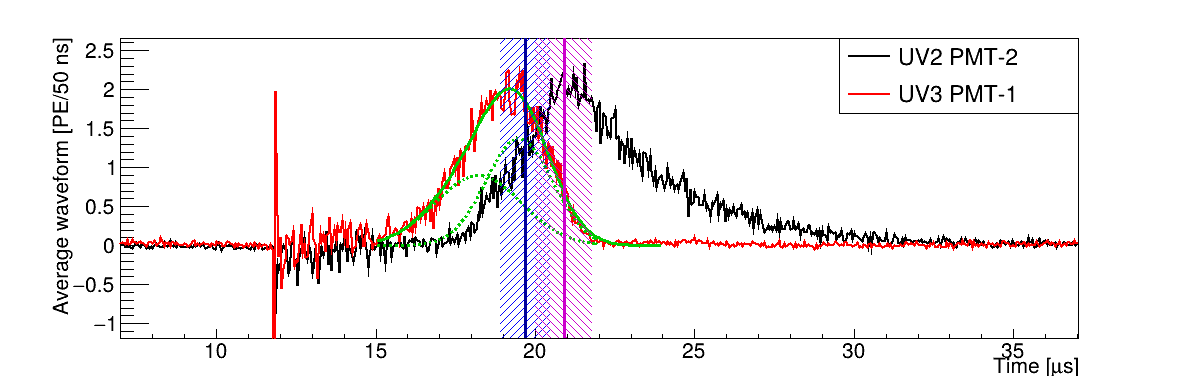}
    \caption{Averaged waveforms in the BF configuration at 3~bar, after subtraction of the corresponding DF signals, for UV2 PMT-2 and UV3 PMT-1. The shaded regions indicate the maximum expected drift time, $7.9 \pm 0.8$~$\upmu$s (blue), and the maximum time to cross the EL region, $1.2 \pm 0.1$~$\upmu$s (purple). The UV3 waveform is fitted with the sum of two Gaussian components (green solid line), with the individual components also shown (green dotted lines).}
    \label{fig:av_waveform_subs3b}
\end{figure}

The averaged waveforms in the BF configuration at 3~bar, after subtraction of the corresponding DF signals, are shown in
Figure~\ref{fig:av_waveform_subs3b} for representative UV2 and UV3 channels.
As in the 1~bar case, a clear difference in temporal structure between the two spectral components is observed. The UV3 signal is faster, whereas the UV2 signal displays a broader time profile. In addition, the UV3 waveform exhibits a more complex temporal structure than that observed at 1~bar. A pronounced delay between the two signals is also visible, with the UV2 emission peaking approximately $2~\upmu$s after the UV3 component.

To describe this behavior phenomenologically, the UV3
signal is fitted with the sum of two Gaussian components, as shown in
Figure~\ref{fig:av_waveform_subs3b}. This fit is not intended to provide a
unique physical model of the emission, but rather to separate a possible early
UV3 contribution from a later component temporally correlated with the UV2
signal.

We estimate the ratio between the photon yields in the UV3 and UV2 spectral
ranges by integrating the corresponding waveforms. After correcting for the
PMT quantum efficiencies and the transmission of the MgF$_2$ windows,
integration of the full UV3 waveform yields a UV3/UV2 ratio of
$20.0 \pm 6.1$\%. Since this value includes both Gaussian components, it does
not represent exclusively the nominal electroluminescence produced by electrons
crossing the central EL gap.

The first Gaussian component peaks around 6~$\mu$s after the trigger and starts
to rise at about 4~$\mu$s. Its origin is not yet understood and may involve
either a different emission mechanism or electroluminescence produced outside
the central EL gap in regions affected by field distortions when both the drift
and EL fields are applied. This early component cannot be unambiguously
associated with the nominal EL emission, since it reaches its maximum before
the expected arrival of the electrons in the central high-field region. Its
physical origin will be investigated in future measurements with a different
experimental setup.

The second Gaussian component starts less than 1~$\mu$s before the UV2 signal
and peaks at a time consistent with the expected maximum electron transit time
across the chamber. This component is therefore interpreted as UV3 emission
associated with electrons reaching and crossing the central EL gap. When the
UV3/UV2 ratio is computed using only this second component, a value of
$10.0 \pm 3.0$\% is obtained, compatible with the result measured at 1~bar.

The pronounced delay observed at 3 bar between the UV3 and UV2 signals is unexpected. At higher pressure, the excimer formation time is expected to be shorter, while the triplet lifetime is not expected to increase sufficiently to account for the larger temporal separation between the two waveforms. Reproducing the UV2 signal from the full UV3 waveform would require modifying the emission response function beyond the model used in Eq.~\eqref{eq:h_2}, for which no complete physical description is presently available.

For this reason, no fit of the UV2 waveform using the full UV3 signal is attempted at 3~bar. Instead, we use the phenomenological decomposition of the UV3 waveform introduced above and test whether the second Gaussian component can account for the observed UV2 emission. This component is compatible with the expected electron transit time across the chamber for the applied fields and can be temporally correlated with the UV2 signal.

To this end, a predicted UV2 waveform is constructed by convolving only this second UV3 Gaussian component with the second-continuum emission response defined in Eq.~\eqref{eq:h_2}. The values of $\tau_f$ and $\tau_t$ are fixed independently, with $\tau_f = 37.6$~ns~\cite{Santorelli:2020fxn} and $\tau_t = 3.12~\upmu\mathrm{s}$ extracted from an exponential fit to the tail of the UV2 signal in this configuration.  A conservative relative uncertainty of $\pm 10\%$ is assigned independently to both parameters.

\begin{figure}[!ht]
    \centering
    \includegraphics[width=\linewidth]{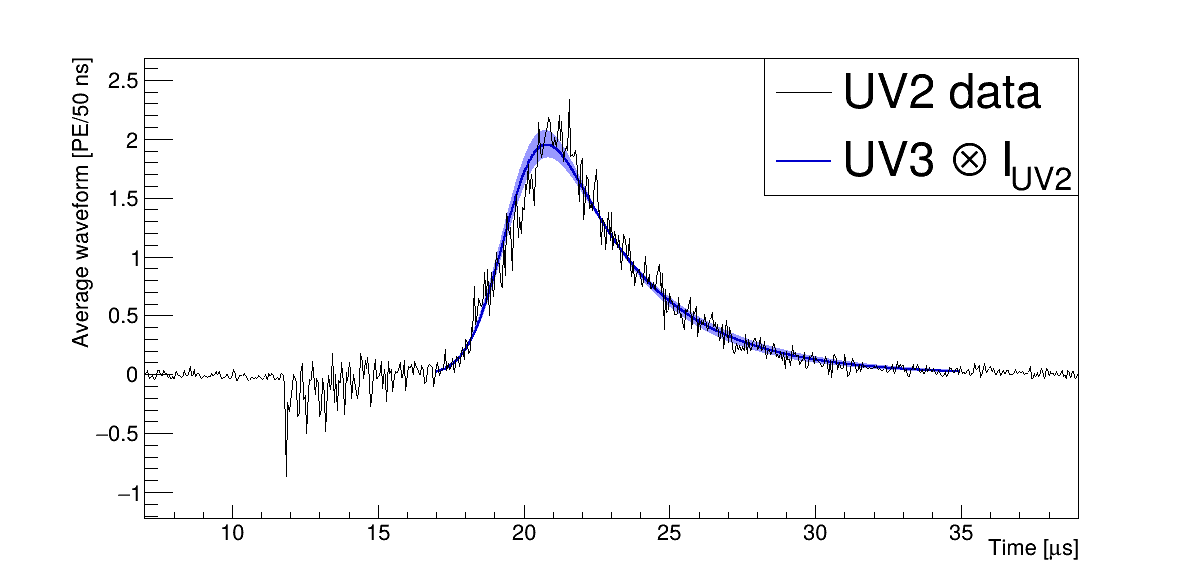}
    \caption{Predicted UV2 waveform obtained by convolving the second Gaussian component of the UV3 signal with the second-continuum emission response defined in Eq.~\eqref{eq:h_2}, compared with the measured UV2 waveform. The physical parameters of the response function are fixed independently, as described in the text.}
    \label{fig:3_bar_initial}
\end{figure}

The result is shown in Figure~\ref{fig:3_bar_initial}, where the predicted UV2 waveform is compared with the measured one. The shaded region represents the uncertainty band obtained from the independent variation of $\tau_f$ and $\tau_t$. The predicted waveform qualitatively reproduces the measured UV2 signal within the quoted uncertainties. This supports the interpretation that the second Gaussian component of the UV3 signal is associated with the regular EL emission, since its convolution with the second-continuum response gives the expected UV2 time profile. 

The nature of the early UV3 component, as well as the larger UV3/UV2 ratio obtained when integrating the full UV3 waveform at 3~bar, remains under investigation. Dedicated measurements with a larger chamber are being prepared to clarify whether this contribution is due to an additional emission mechanism or to instrumental effects related to the present field configuration.

%% file: Sec/Conclusions.tex
\section{Conclusions and implications for electroluminescence TPCs}
\label{sec:Concl}

In this work, we have presented a wavelength-resolved study of
electroluminescence light in gaseous argon using a compact TPC equipped with
PMTs sensitive to two complementary spectral regions: the UV2 range, covering
approximately 110--160~nm, and the UV3 range, extending from 160~nm to the
UV--visible region. This study is directly relevant to rare-event searches
based on dual-phase noble-element TPCs, where electroluminescence is used to
amplify small ionization signals and where the optical response plays a central
role in energy reconstruction, position reconstruction, and background
rejection.

The results show that argon electroluminescence cannot be described as a
quasi-monochromatic emission at 128~nm. In addition to the dominant VUV
second-continuum emission detected in the UV2 region, a clear contribution is
observed in the UV3 spectral range. After correcting for PMT quantum efficiency
and MgF$_2$ window transmission, the UV3 contribution at 1~bar is measured to
be $9.8 \pm 2.9$\% relative to the UV2 emission. Longer-wavelength components
therefore represent a non-negligible contribution to the EL signal and should
be included in optical models of argon-based detectors.

More importantly, the two spectral components carry different temporal
information. The fast UV3 emission closely follows the electrons crossing the
high-field region and therefore provides a direct timing observable of their
transit through the EL gap. In contrast, the UV2 signal is broadened by the
formation and de-excitation of argon excimers responsible for the second
continuum. At 1~bar, the UV2 waveform can be reproduced by convolving the
measured UV3 signal with the second-continuum emission response. This supports
a phenomenological interpretation in which the UV3 signal traces the underlying
charge transport with substantially less temporal smearing, while the UV2
signal results from the subsequent excimer formation and decay.

A preliminary study at 3~bar was also performed under comparable reduced-field
conditions. At this pressure, the UV3/UV2 ratio obtained by integrating the full
waveform reaches $20.0 \pm 6.1$\%. However, the UV3 waveform exhibits a more
complex temporal structure, with a pronounced delay between the UV3 and UV2
signals. When the UV3 integral is restricted to the second Gaussian component,
which is temporally correlated with the UV2 signal and consistent with the
expected electron transit through the chamber, the ratio decreases to
$10.0 \pm 3.0$\%, in agreement with the value measured at 1~bar. The operating
conditions and the measured UV3/UV2 ratios at the two pressures are summarized
in Table~\ref{tab:el_summary}.

\begin{table*}[t]
\centering
\setlength{\tabcolsep}{8pt}
\caption{Summary of the electroluminescence measurements performed at
different argon pressures. The value reported on the second line at 3~bar is
obtained by restricting the UV3 integral to the second Gaussian component of
the phenomenological fit, which is temporally correlated with the UV2 signal
and is consistent with the expected electron transit time across the chamber.}
\begin{tabular}{cccccc}
\hline
Pressure &
$E_{\mathrm{drift}}$ &
$E_{\mathrm{EL}}$ &
$E_{\mathrm{drift}}/P$ &
$E_{\mathrm{EL}}/P$ &
UV3/UV2 \\
(bar) &
($\mathrm{kV\,cm^{-1}}$) &
($\mathrm{kV\,cm^{-1}}$) &
($\mathrm{kV\,cm^{-1}\,bar^{-1}}$) &
($\mathrm{kV\,cm^{-1}\,bar^{-1}}$) &
(\%) \\
\hline
$1.08 \pm 0.05$ &
$0.266 \pm 0.014$ &
$1.69 \pm 0.09$ &
$0.246 \pm 0.018$ &
$1.56 \pm 0.11$ &
$9.8 \pm 2.9$ \\[3pt]

$3.00 \pm 0.15$ &
$0.738 \pm 0.038$ &
$4.72 \pm 0.24$ &
$0.246 \pm 0.018$ &
$1.57 \pm 0.11$ &
$20.0 \pm 6.1$ \\

& & & & & $(10.0 \pm 3.0)$ \\
\hline
\end{tabular}
\label{tab:el_summary}
\end{table*}

The different temporal responses of the two spectral components give
wavelength-resolved detection a practical role beyond photon counting. In an
electroluminescence TPC, the UV3 channel could improve the reconstruction of
the longitudinal charge distribution and help identify overlapping or piled-up
S2 signals that would otherwise be merged by the slower UV2 response.

Wavelength-resolved detection may also help separate primary scintillation from
secondary electroluminescence when the two signals overlap in time. Because S1
and S2 light are produced through different processes and exhibit different
spectral and temporal compositions, a fast UV3-sensitive channel could reveal a
small primary-scintillation contribution in a much larger S2
pulse. This capability may be particularly relevant for accidental coincidences
between S1-only events and low-energy S2-only-like events near the liquid-argon
surface, as well as for other event topologies containing partially overlapping
light signals. These observables are not currently exploited in standard
noble-gas detectors, where light is generally integrated over broad wavelength
ranges.

Taken together with our previous measurements of particle-induced
scintillation~\cite{Santorelli:2020fxn}, the present results show that
wavelength-resolved detection is sensitive to the process by which argon light
is produced. The results reported in our previous study suggest that the prompt
UV3 contribution is more pronounced for $\alpha$-induced scintillation than
for $\beta$-induced scintillation. These observations strengthen the
experimental basis for the spectroscopy-based particle-identification strategy
first proposed in our previous work. New measurements of $\alpha$- and
$\beta$-induced scintillation with the same wavelength-sensitive setup are
ongoing to quantify the achievable discrimination.

A more thorough characterization was not possible with the present experimental
setup. Dedicated measurements with an upgraded and larger wavelength-sensitive
detector are being prepared to study argon electroluminescence under conditions
closer to those of large-scale dual-phase TPCs. These measurements will be
needed to quantify the achievable improvement in electron-transit-time
reconstruction, pile-up rejection, S1--S2 separation, and particle
discrimination.

%% file: Sec/Acknowledgments.tex
\section*{Acknowledgments}
\label{sec:Acknow}

This work was made possible by funding from the Spanish Ministry of Science and Innovation (MICINN) under Grants PID2022-138357NB-C22 and EUR2023-143480 (Europa Excelencia).

%% file: bibliografia.bib
@article{Santorelli:2020fxn,
  author  = {Santorelli, R. and Sanchez Garcia, E. and Abia, P. Garcia and Gonz\'alez-D\'\i{}az, D. and Manzano, R. Lopez and Morales, J. J. Martinez and Pesudo, V. and Romero, L.},
  title   = {Spectroscopic analysis of the gaseous argon scintillation with a wavelength sensitive particle detector},
  journal = {Eur. Phys. J. C},
  volume  = {81},
  number  = {7},
  pages   = {622},
  year    = {2021},
  doi     = {10.1140/epjc/s10052-021-09375-3}
}

@article{Leardini:2021qnf,
  author  = {Leardini, S. and S\'anchez Garc\'\i{}a, E. and Amedo, P. and Saa-Hern\'andez, A. and Gonz\'alez-D\'\i{}az, D. and Santorelli, R. and Fern\'andez-Posada, D. J. and Gonz\'alez, D.},
  title   = {Time and band-resolved scintillation in time projection chambers based on gaseous xenon},
  journal = {Eur. Phys. J. C},
  volume  = {82},
  number  = {5},
  pages   = {425},
  year    = {2022},
  doi     = {10.1140/epjc/s10052-022-10385-y}
}

@article{ArDM:2017ndf,
  author        = {Calvo, J. and others},
  collaboration = {ArDM},
  title         = {Backgrounds and pulse shape discrimination in the ArDM liquid argon TPC},
  journal       = {JCAP},
  volume        = {2018},
  number        = {12},
  pages         = {011},
  year          = {2018},
  doi           = {10.1088/1475-7516/2018/12/011}
}

@article{DarkSide:2018bpj,
  author        = {Agnes, P. and others},
  collaboration = {DarkSide},
  title         = {Low-Mass Dark Matter Search with the DarkSide-50 Experiment},
  journal       = {Phys. Rev. Lett.},
  volume        = {121},
  number        = {8},
  pages         = {081307},
  year          = {2018},
  doi           = {10.1103/PhysRevLett.121.081307}
}

@article{XENON:2024ijk,
  author        = {Aprile, Elena and others},
  collaboration = {XENON},
  title         = {First Indication of Solar {$^{8}$B} Neutrinos via Coherent Elastic Neutrino-Nucleus Scattering with XENONnT},
  journal       = {Phys. Rev. Lett.},
  volume        = {133},
  number        = {19},
  pages         = {191002},
  year          = {2024},
  doi           = {10.1103/PhysRevLett.133.191002}
}

@article{DarkSide:2018ppu,
  author        = {Agnes, P. and others},
  collaboration = {DarkSide},
  title         = {Constraints on Sub-GeV Dark-Matter--Electron Scattering from the DarkSide-50 Experiment},
  journal       = {Phys. Rev. Lett.},
  volume        = {121},
  number        = {11},
  pages         = {111303},
  year          = {2018},
  doi           = {10.1103/PhysRevLett.121.111303}
}

@article{nakamura1988electron,
  author  = {Nakamura, Y. and Kurachi, M.},
  title   = {Electron transport parameters in argon and its momentum transfer cross section},
  journal = {J. Phys. D: Appl. Phys.},
  volume  = {21},
  number  = {5},
  pages   = {718--723},
  year    = {1988},
  doi     = {10.1088/0022-3727/21/5/008}
}

@misc{MATLAB,
  author       = {{The MathWorks, Inc.}},
  title        = {{MATLAB}},
  year         = {2023},
  note         = {Release R2023b}
}

@article{Monteiro2008,
  author  = {Monteiro, C. M. B. and Lopes, J. A. M. and Veloso, J. F. C. A. and dos Santos, J. M. F.},
  title   = {Secondary scintillation yield in pure argon},
  journal = {Physics Letters B},
  volume  = {668},
  pages   = {167--170},
  year    = {2008},
  doi     = {10.1016/j.physletb.2008.08.030}
}
